\documentclass{article}

\usepackage{arxiv}

\usepackage[utf8]{inputenc} 
\usepackage[T1]{fontenc}    
\usepackage{hyperref}       
\usepackage{url}            
\usepackage{booktabs}       
\usepackage{amsfonts}       
\usepackage{nicefrac}       
\usepackage{microtype}      
\usepackage{lipsum}		
\usepackage{graphicx}
\usepackage[numbers]{natbib}
\usepackage{doi}
\usepackage{float}

\usepackage{pifont}

\usepackage{listings}

\usepackage{xcolor}
\definecolor{codegreen}{rgb}{0,0.6,0}
\definecolor{codegray}{rgb}{0.5,0.5,0.5}
\definecolor{codepurple}{rgb}{0.58,0,0.82}
\definecolor{backcolour}{rgb}{1,1,1}
\lstdefinestyle{mystyle}{
    backgroundcolor=\color{backcolour},   
    commentstyle=\color{codegreen},
    keywordstyle=\color{magenta},
    numberstyle=\tiny\color{codegray},
    stringstyle=\color{codepurple},
    basicstyle=\ttfamily\footnotesize,
    breakatwhitespace=false,         
    breaklines=true,                 
    captionpos=t, 
    keepspaces=true,                 
    numbers=none, 
    numbersep=5pt,                  
    showspaces=false,                
    showstringspaces=false,
    showtabs=false,                  
    tabsize=2
}
\makeatletter
\pretocmd\lst@makecaption{\noindent{\rule{\linewidth}{1pt}}}{}{}
\makeatother

\usepackage{rotating}

\title{LightPHE: Integrating Partially Homomorphic Encryption into Python with Extensive Cloud Environment Evaluations}

\author{ \href{https://orcid.org/0000-0002-0345-0088}{\includegraphics[scale=0.06]{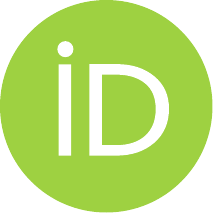}\hspace{1mm}Sefik Ilkin Serengil} \\
	Solution Engineering Department\\
	Vorboss Limited\\
	\texttt{sefik.serengil@vorboss.com} \\
	\And
	\href{https://orcid.org/0000-0003-1250-5949}{\includegraphics[scale=0.06]{orcid.pdf}\hspace{1mm}Alper Ozpinar} \\
	Mechatronics Engineering Department\\
	Istanbul Ticaret University\\
	\texttt{alper.ozpinar@ticaret.edu.tr} \\
}

\renewcommand{\shorttitle}{LightPHE}

\hypersetup{
    pdftitle={LightPHE: A Hybrid Partially Homomorphic Encryption Library for Python},
    pdfsubject={cs.CR},
    pdfauthor={Author 1, Author 2},
    pdfkeywords={Cryptography, Homomorphic Encryption, Cloud Benchmarks},
}

\begin{document}
\maketitle

\begin{abstract}
Homomorphic encryption facilitates computations on encrypted data without the need to access private keys, enabling secure updates of sensitive information directly within cloud environments. Absent this technology, updates must be performed on-premises, or private keys must be transmitted to the cloud, thereby increasing security vulnerabilities. While fully homomorphic encryption (FHE) supports both additive and multiplicative operations on ciphertexts, partially homomorphic encryption (PHE) supports either addition or multiplication, providing a more efficient and practical solution for various applications.

In this study, LightPHE, a lightweight and hybrid PHE framework designed for Python, is introduced to address the paucity of existing PHE libraries. LightPHE integrates multiple PHE algorithms, adhering to best practices in software engineering to ensure robustness and usability. The framework offers a modular and extensible design, facilitating rapid prototyping and development of secure applications.

Comprehensive cloud-based experiments were conducted to evaluate the performance and scalability of LightPHE across diverse cloud environments, including Google Colab (Normal, A100 GPU, L4 GPU, T4 High RAM, TPU2) and Microsoft Azure Spark. These experiments focused on key metrics such as key generation, encryption, decryption, and homomorphic operations. The results highlighted LightPHE's efficiency, demonstrating superior performance in high-computation environments like Colab A100 GPU and TPU2, while also providing viable options for cost-effective cloud setups like Colab Normal and Azure Spark.

Comparative analyses revealed that LightPHE performs optimally in terms of computational efficiency and scalability, making it suitable for various practical applications. The benchmarks provided valuable insights into the selection of appropriate cloud environments based on specific use cases and performance requirements. This study underscores LightPHE's potential to advance the implementation of homomorphic encryption in diverse domains, promoting secure and efficient data processing in the cloud.
\end{abstract}


\section{Introduction}

In today's digital age, the protection of sensitive information has become more critical than ever. Homomorphic encryption enables calculations on encrypted data without the need to access private keys \cite{first_homomoprhic_encryption}. This capability allows data such as salaries and pensions to be stored in an encrypted form and securely updated without decryption. This approach has gained significant popularity with the proliferation of cloud computing. Companies increasingly store their data in the cloud rather than on-premise systems, elevating the importance of robust security measures.

Without homomorphic encryption, encrypted data must be decrypted, updated, and then re-encrypted. All updates must be performed on-premises, or the private key must be transmitted to the cloud, which introduces security risks. With homomorphic encryption, sensitive data can be stored and updated directly in the cloud without exposing the private key.

Fully homomorphic encryption (FHE) allows for both addition and multiplication on ciphertexts \cite{fhe}. Conversely, partially homomorphic encryption (PHE) is more limited, supporting either addition or multiplication, but not both \cite{phe}. PHE algorithms can also multiply a ciphertext by a known constant and support regenerating ciphertexts, meaning the same plaintext can have different ciphertext representations. PHE schemes find applications in various contexts such as e-voting and Private Information Retrieval \cite{he_survey}. However, these applications have been limited in scope due to constraints on the types of homomorphic evaluation operations they support. Essentially, PHE schemes are only applicable to specific scenarios where the algorithms involve either addition or multiplication operations exclusively.

Although FHE has become more accessible, PHE is often a more efficient and practical choice for many tasks. PHE is faster, requires fewer computational resources, generates smaller ciphertexts, and uses smaller keys. It is well-suited for environments with limited memory and strikes a good balance for practical applications.

In this paper, LightPHE, a lightweight and hybrid partially homomorphic encryption framework for Python, is proposed. It is open-sourced at \url{https://github.com/serengil/lightphe} and licensed under MIT. While several FHE options are available, such as SEAL \cite{seal}, TenSEAL \cite{tenseal}, Pyfhel \cite{pyfhel}, or PyFHE \cite{pyfhe}, there is a notable lack of PHE frameworks. LightPHE aims to address this gap by providing a library that encapsulates various PHE algorithms, including RSA \cite{rsa}, ElGamal \cite{elgamal}, Exponential ElGamal, Elliptic Curve ElGamal \cite{ecelgamal}, Paillier \cite{paillier}, Damgard-Jurik \cite{damgard}, Okamoto–Uchiyama \cite{okamoto}, Benaloh \cite{benaloh}, Naccache–Stern \cite{naccache}, and Goldwasser–Micali \cite{golwasser}. The design of the framework adheres to best practices in software engineering.

In Section \ref{sec:algorithms}, the various algorithms supported by LightPHE are explored. Each algorithm's capabilities and the theoretical underpinnings of their homomorphic features are explained. A range of algorithms, including RSA, ElGamal, Paillier, and others, are covered. The application of these algorithms to encrypted data, ensuring the security of sensitive information without the need for decryption, is demonstrated.

Moving on to Section \ref{sec:software_architecture}, a detailed examination of the design and structure of LightPHE is provided. The different components of the framework are broken down and their integration is explained. Using diagrams and clear explanations, the management of tasks such as encryption, computation, and decryption by LightPHE is illustrated. This section aims to enhance readers' understanding of the framework's functionality and reliability.

In Section \ref{sec:implementation}, the practical use of LightPHE is detailed. Examples and code snippets are provided to guide users through integrating LightPHE into their Python projects. By breaking down the implementation details and offering practical advice, the section aims to facilitate researchers and developers in maximizing the utilization of LightPHE's features.

In addition to proposing LightPHE, extensive experiments were conducted to evaluate its performance across various cloud environments, including Colab Normal, Colab A100 GPU, Colab L4 GPU, Colab T4 High RAM, Colab TPU2, and Azure Spark. These experiments focused on several critical parameters, including time consumption for key generation, encryption, decryption, and homomorphic operations such as addition or multiplication for different key sizes.

Recent advancements in cloud computing and homomorphic encryption have significantly influenced the design and deployment of secure and efficient cryptographic systems. By leveraging the unique capabilities of these cloud environments, detailed benchmarks were established to provide a comprehensive performance analysis of LightPHE. This analysis not only highlights the practical differences between various cloud setups but also serves as a valuable benchmark for practitioners and researchers in selecting the appropriate environment for their specific needs.

Radar maps were created to visualize and compare the performance metrics across these diverse cloud environments. These visual tools offer insights into the strengths and weaknesses of each setup, aiding in the decision-making process for deploying homomorphic encryption in production systems. The comparative analysis emphasizes factors such as computational efficiency, scalability, and resource utilization, which are crucial for determining the suitability of a particular cloud environment for real-world applications.

Furthermore, the benchmarks provide guidance on selecting the optimal cloud environment based on specific use cases and performance requirements. For instance, environments like Colab A100 GPU and TPU2 are evaluated for their superior computational power, making them suitable for tasks requiring high performance and scalability. Conversely, setups like Colab Normal and Azure Spark are assessed for their cost-effectiveness and accessibility, offering viable options for projects with limited resources.

The insights gained from these experiments are intended to assist organizations in making informed decisions about the deployment of homomorphic encryption technologies in their production systems. By understanding the performance characteristics of different cloud environments, stakeholders can better align their infrastructure choices with their operational and security requirements, ensuring a robust and efficient implementation of LightPHE.

\section{Algorithms}
\label{sec:algorithms}

LightPHE wraps RSA, ElGamal, Exponential ElGamal, Elliptic Curve ElGamal, Paillier, Damgard-Jurik, Okamoto–Uchiyama, Benaloh, Naccache–Stern, and Goldwasser–Micali algorithms. Table~\ref{tab:algorithms} shows each algorithms homomorphic fetures. In this section, proofs of their homomorphic features will be covered.

\begin{minipage}{\linewidth} 
\begin{table}[H]
\caption{Supported Algorithms in LightPHE}
\centering
\begin{tabular}{lllllll}
\toprule
Algorithm              & Year & \begin{tabular}[c]{@{}l@{}}Homomorphic\\ Multiplication\end{tabular} & \begin{tabular}[c]{@{}l@{}}Homomorphic\\ Addition\end{tabular} & \begin{tabular}[c]{@{}l@{}}Scalar\\ Multiplication\end{tabular} & \begin{tabular}[c]{@{}l@{}}Homomorphic\\ XOR\end{tabular} & \begin{tabular}[c]{@{}l@{}}Ciphertext\\ Regeneration\end{tabular} \\
\midrule
RSA & 1977 & $\checkmark$ & × & × & × & × \\
Goldwasser-Micali & 1982 & × & × & × & $\checkmark$ & × \\
ElGamal & 1985 & $\checkmark$ & × & × & × & × \\
Exp. ElGamal & 1985 & × & $\checkmark$ & $\checkmark$ & × & $\checkmark$ \\
Benaloh & 1985 & × & $\checkmark$ & $\checkmark$ & × & $\checkmark$ \\
EC ElGamal & 1998 & × & $\checkmark$ & $\checkmark$ & × & × \\
Naccache-Stern & 1998 & × & $\checkmark$ & $\checkmark$ & × & $\checkmark$ \\
Okamoto-Uchiyama & 1998 & × & $\checkmark$ & $\checkmark$ & × & $\checkmark$ \\
Paillier & 1999 & × & $\checkmark$ & $\checkmark$ & × & $\checkmark$ \\
Damgard-Jurik & 2001 & × & $\checkmark$ & $\checkmark$ & × & $\checkmark$ \\
\bottomrule
\end{tabular}
\label{tab:algorithms}
\end{table}
\end{minipage}
\newline
\newline


\subsection{RSA}
\label{sec:rsa}

RSA invented in 1977 by Ron Rivest, Adi Shamir and Leonard Adleman \cite{rsa}. It is depending on the difficulty of factoring large integers. The cryptosystem show multiplicatively homomorphic features. RSA encryption requires raising the message to the power of the public key e and then taking the result modulo n as show in Formula~\ref{eq:rsa_generic}.

\begin{equation}
\label{eq:rsa_generic}
    {\varepsilon(m)} = { (m)^e \quad mod \quad n }
\end{equation}

If you encrypt a plaintext pair $m_1$ and $m_2$ with RSA, then their corresponding ciphertexts will be calculated using Equations \ref{eq:rsa_c1} and \ref{eq:rsa_c2}.

\begin{equation}
\label{eq:rsa_c1}
    {\varepsilon(m_1)} = { (m_1)^e \quad mod \quad n }
\end{equation}

\begin{equation}
\label{eq:rsa_c2}
    {\varepsilon(m_2)} = { (m_2)^e \quad mod \quad n }
\end{equation}

The multiplication of encrypted values will be calculated using Equation \ref{eq:rsa_c1_times_c2}.

\begin{equation}
\label{eq:rsa_c1_times_c2}
    {\varepsilon(m_1)} \times {\varepsilon(m_2)} = { (m_1)^e(m_2)^e \quad mod \quad n }
\end{equation}

Multiplication of ciphertexts can be reorganized as shown in Equation \ref{eq:rsa_c1_times_c2_organized}.

\begin{equation}
\label{eq:rsa_c1_times_c2_organized}
    {\varepsilon(m_1)} \times {\varepsilon(m_2)} = { (m_1 \times m_2)^e \quad mod \quad n }
\end{equation}

On the other hand, encryption of the multiplication of plain $m_1$ and $m_2$ will be same as shown in Equation \ref{eq:m1_times_m2_encrypted}.

\begin{equation}
\label{eq:m1_times_m2_encrypted}
    {\varepsilon(m_1 \times m_2)} = { (m_1 \times m_2)^e \quad mod \quad n }
\end{equation}

In summary, RSA is homomorphic with respect to the multiplication as shown in Equation \ref{eq:rsa_summary}.

\begin{equation}
\label{eq:rsa_summary}
    {\varepsilon(m_1 \times m_2)} = {\varepsilon(m_1)} \times {\varepsilon(m_2)}
\end{equation}


\subsection{ElGamal}
\label{sec:elgamal}

ElGamal algorithm was invented by Taher Elgamal in 1985 \cite{elgamal}. It is depending on the difficulty of computing discrete logarithms over finite fields. The algorithm shows multiplicatively homomorphic features. ElGamal encryption requires to calculate of tuple of a public generator to the power of a random integer and plaintext times public key to the power of same random key as shown in Equation \ref{eq:elgamal_generic}.

\begin{equation}
\label{eq:elgamal_generic}
    {\varepsilon(m, r)} = ({ g^r, m \times h^r }) \quad mod \quad p
\end{equation}

If you encrypt a plaintext pair $m_1$ and $m_2$ with ElGamal, then their corresponding ciphertexts will be calculated using Equations \ref{eq:elgamal_c1} and \ref{eq:elgamal_c2}.

\begin{equation}
\label{eq:elgamal_c1}
    {\varepsilon(m_1, r_1)} = ({ g^{r_1}, m_1 \times h^{r_1} }) \quad mod \quad p
\end{equation}

\begin{equation}
\label{eq:elgamal_c2}
    {\varepsilon(m_2, r_2)} = ({ g^{r_2}, m_2 \times h^{r_2} }) \quad mod \quad p
\end{equation}

The multiplication of encrypted values will be calculated using Equation \ref{eq:elgamal_c1_times_c2}.

\begin{equation}
\label{eq:elgamal_c1_times_c2}
    {\varepsilon(m_1, r_1)} \times {\varepsilon(m_2, r_2)} = ({ g^{r_1}, m_1 \times h^{r_1} }) \times ({ g^{r_2}, m_2 \times h^{r_2} }) \quad mod \quad p
\end{equation}

Multiplication of ciphertexts can be reorganized as shown in Equation \ref{eq:elgamal_c1_times_c2_organized} and \ref{eq:elgamal_c1_times_c2_organized_2}.

\begin{equation}
\label{eq:elgamal_c1_times_c2_organized}
    {\varepsilon(m_1, r_1)} \times {\varepsilon(m_2, r_2)} = ({ g^{r_1} \times g^{r_2}, m_1 \times m_2 \times h^{r_1} } \times h^{r_2}) \quad mod \quad p
\end{equation}

\begin{equation}
\label{eq:elgamal_c1_times_c2_organized_2}
    {\varepsilon(m_1, r_1)} \times {\varepsilon(m_2, r_2)} = ({ g^{r_1 + r_2}, m_1 \times m_2 \times h^{r_1 + r_2} }) \quad mod \quad p
\end{equation}

On the other hand, multiplication of plain $m_1$ and $m_2$ with random key $r_1$+$r_2$ will give same result as shown in Equation \ref{eq:elgamal_c1_times_c2_organized_3}.

\begin{equation}
\label{eq:elgamal_c1_times_c2_organized_3}
    {\varepsilon(m_1 \times m_2, r_1 + r_2)} = ({ g^{r_1 + r_2}, m_1 \times m_2 \times h^{r_1 + r_2} }) \quad mod \quad p
\end{equation}

In conclusion, ElGamal is homomorphic with respect to the multiplication as shown in Equation \ref{eq:elgamal_summary}.

\begin{equation}
\label{eq:elgamal_summary}
    {\varepsilon(m_1, r_1)} \times {\varepsilon(m_2, r_2)} = {\varepsilon(m_1 \times m_2, r_1 + r_2)}
\end{equation}


\subsection{Exponential ElGamal}
\label{sec:exponential_elgamal}

ElGamal encryption requires to calculate a tuple and the second item of the tuple has plaintext m as multiplier. If this item is modified to generator g to the power of plaintext m, then the algorithm will start to show additive homomorphic features but it will lose its multiplicative homomorphic features.

\begin{equation}
\label{eq:exponential_elgamal_generic}
    {\varepsilon(m, r)} = ({ g^r, g^m \times h^r }) \quad mod \quad p
\end{equation}

For instance, encryption of plaintext $m_1$ and $m_2$ couple with random keys $r_1$ and $r_2$ will be calculated using Equations \ref{eq:exponential_elgamal_c1} and \ref{eq:exponential_elgamal_c2}.

\begin{equation}
\label{eq:exponential_elgamal_c1}
    {\varepsilon(m_1, r_1)} = ({ g^{r_1}, g^{m_1} \times h^{r_1} }) \quad mod \quad p
\end{equation}

\begin{equation}
\label{eq:exponential_elgamal_c2}
    {\varepsilon(m_2, r_2)} = ({ g^{r_2}, g^{m_2} \times h^{r_2} }) \quad mod \quad p
\end{equation}

Thereafter, multiplication of ciphertexts will be calculated using Equation \ref{eq:exponential_elgamal_c1_times_c2}.

\begin{equation}
\label{eq:exponential_elgamal_c1_times_c2}
    {\varepsilon(m_1, r_1)} \times {\varepsilon(m_2, r_2)} = ({ g^{r_1}, g^{m_1} \times h^{r_1} }) \times ({ g^{r_2}, g^{m_2} \times h^{r_2} }) \quad mod \quad p
\end{equation}

That multiplication can be reorganized as shown in Equations \ref{eq:exponential_elgamal_c1_times_c2_reorganized} and \ref{eq:exponential_elgamal_c1_times_c2_reorganized_2}.

\begin{equation}
\label{eq:exponential_elgamal_c1_times_c2_reorganized}
    {\varepsilon(m_1, r_1)} \times {\varepsilon(m_2, r_2)} 
    = 
    (
        { g^{r_1} \times g^{r_2},
        g^{m_1} \times g^{m_2} \times h^{r_1} \times h^{r_2} }
    ) \quad mod \quad p
\end{equation}

\begin{equation}
\label{eq:exponential_elgamal_c1_times_c2_reorganized_2}
    {\varepsilon(m_1, r_1)} \times {\varepsilon(m_2, r_2)} 
    = 
    (
        { g^{r_1 + r_2},
        g^{m_1 + m_2} \times h^{r_1 + r_2} }
    ) \quad mod \quad p
\end{equation}

On the other hand, encryption of the addition of $m_1$ and $m_2$ with random key $r_1$ + $r_2$ will give same results as shown in Equation \ref{eq:exponential_elgamal_encryption_m1_plus_m2}.

\begin{equation}
\label{eq:exponential_elgamal_encryption_m1_plus_m2}
    {\varepsilon(m_1 + m_2, r_1 + r_2)}
    = 
    (
        { g^{r_1 + r_2},
        g^{m_1 + m_2} \times h^{r_1 + r_2} }
    ) \quad mod \quad p
\end{equation}

In summary, exponential ElGamal is homomorphic with respect to the addition as shown in Equation \ref{eq:exponential_elgamal_summary}. However, it is not homomorphic with respect to the multiplication similar to standard ElGamal.

\begin{equation}
\label{eq:exponential_elgamal_summary}
    {\varepsilon(m_1, r_1)} \times {\varepsilon(m_2, r_2)} 
    =
    {\varepsilon(m_1 + m_2, r_1 + r_2)}
\end{equation}

However, exponential ElGamal comes with some limitations. Once a ciphertext is decrypted, it will give $g^m$ instead of plaintext $m$. To restore m from $g^m$, discrete logarithm problem must be resolved which is hard for large integers. This makes exponential ElGamal theoritical algorithm instead of practical one. On the other hand, if the plaintext is millonish value, then current processing powers will be able to find it very fast. So, the algorithm can be adopted according to the size of plaintext.

\subsubsection{Ciphertext Regeneration Feature}

Exponential ElGamal is additively homomorphic and the neutral element in addition is 0. Encrypting the neutral element and then add it into a ciphertext will not change the corresponding plaintext but ciphertext will have a different representation. In that way, ciphertext can be regenerated as shown in Equations \ref{eq:exponential_elgamal_neutral_1} and \ref{eq:exponential_elgamal_neutral_2}where \textit{D} denotes decryption.

\begin{equation}
\label{eq:exponential_elgamal_neutral_1}
    {\varepsilon(m_1, r_1)} \times {\varepsilon(0, r_2)}
    \neq
    {\varepsilon(m_1, r_1)}
\end{equation}

\begin{equation}
\label{eq:exponential_elgamal_neutral_2}
    D({\varepsilon(m_1, r_1)} \times {\varepsilon(0, r_2)} )
    =
    D({\varepsilon(m_1, r_1)})
\end{equation}

\subsubsection{Scalar Multiplication Feature}

Finding the k-th power of a ciphertext is shown in Equation \ref{eq:exponential_elgamal_scalar_1} and \ref{eq:exponential_elgamal_scalar_2}.

\begin{equation}
\label{eq:exponential_elgamal_scalar_1}
    {\varepsilon(m_1, r_1)}^k = ({ g^{r_1}, g^{m_1} \times h^{r_1} }) ^k \quad mod \quad p
\end{equation}

\begin{equation}
\label{eq:exponential_elgamal_scalar_2}
    {\varepsilon(m_1, r_1)}^k = ({ g^{r_1 \times k}, g^{m_1 \times k} \times h^{r_1 \times k} }) \quad mod \quad p
\end{equation}

On the other hand, scalar multiplication of k and $m_1$ with random key k times $r_1$ will give the same result as shown in Equation \ref{eq:exponential_elgamal_scalar_3}. This proves the scalar multiplication feature of exponential ElGamal.

\begin{equation}
\label{eq:exponential_elgamal_scalar_3}
    {\varepsilon(m_1 \times k, r_1 \times k)} = ({ g^{r_1 \times k}, g^{m_1 \times k} \times h^{r_1 \times k} }) \quad mod \quad p
\end{equation}

After decryption, only the first parameter of the encryption function will be restored, representing the plaintext multiplied by a constant value.


\subsection{Elliptic Curve ElGamal}
\label{sec:ec_elgamal}

Co-usage of elliptic curve cryptography and ElGamal algorithm offers additively homomorphic features \cite{ecelgamal}. Elliptic curve cryptography comes with much smaller key lengths with same level accuracy when compared to RSA or ElGamal. The most common elliptic curve forms are Weierstrass \cite{weierstrass}, Koblitz \cite{koblitz} and Edwards \cite{edwards}. The elliptic curve forms differentiate addition formulas but concept of homomorphy becomes same. LightPHE just covers elliptic curves in Weierstrass form.

Similar to ElGamal algorithm, Elliptic Curve ElGamal creates tuples in the ciphertext, but each item of the tuple is also a tuple of 2-dimensional point with x and y coordinates. A generalized encryption procedure of Elliptic Curve ElGamal is shown in Equation\ref{eq:ecelgamal_generic} where \textit{G} is the public base point and \textit{Q} is the public key point calculated by private key times base point \textit{G}.

\begin{equation}
\label{eq:ecelgamal_generic}
    {\varepsilon(m, r)} = (r \times G, r \times Q + m \times G)
\end{equation}

Then, encryption of plaintext pairs m1 and m2 with random keys r1 and r2 will be calculated using Equations \ref{eq:ecelgamal_c1} and \ref{eq:ecelgamal_c2}.

\begin{equation}
\label{eq:ecelgamal_c1}
    {\varepsilon(m_1, r_1)} = (r_1 \times G, r_1 \times Q + m_1 \times G)
\end{equation}

\begin{equation}
\label{eq:ecelgamal_c2}
    {\varepsilon(m_2, r_2)} = (r_2 \times G, r_2 \times Q + m_2 \times G)
\end{equation}

Thereafter, addition of ciphertexts will be calculated using Equation \ref{eq:ecelgamal_c1_times_c2}.

\begin{equation}
\label{eq:ecelgamal_c1_times_c2}
    {\varepsilon(m_1, r_1)} + {\varepsilon(m_2, r_2)} = (r_1 \times G, r_1 \times Q + m_1 \times G) + (r_2 \times G, r_2 \times Q + m_2 \times G)
\end{equation}

The addition can be reorganized as shown in Equations \ref{eq:ecelgamal_c1_times_c2_reorganized} and \ref{eq:ecelgamal_c1_times_c2_reorganized_2}.

\begin{equation}
\label{eq:ecelgamal_c1_times_c2_reorganized}
    {\varepsilon(m_1, r_1)} + {\varepsilon(m_2, r_2)} = (r_1 \times G + r_2 \times G, r_1 \times Q + m_1 \times G + r_2 \times Q + m_2 \times G)
\end{equation}

\begin{equation}
\label{eq:ecelgamal_c1_times_c2_reorganized_2}
    {\varepsilon(m_1, r_1)} + {\varepsilon(m_2, r_2)} = ((r_1 + r_2) \times G, (r_1 + r_2) \times Q + (m_1 + m_2) \times G)
\end{equation}

On the other hand, encryption of addition of plaintexts $m_1$ and $m_2$ with random key $r_1$ + $r_2$ will give same result as shown in Equation \ref{eq:ecelgamal_encryption_of_m1_plus_m2}.

\begin{equation}
\label{eq:ecelgamal_encryption_of_m1_plus_m2}
    {\varepsilon(m_1 + m_2, r_1 + r_2)} = ((r_1 + r_2) \times G, (r_1 + r_2) \times Q + (m_1 + m_2) \times G)
\end{equation}

In conclusion, Elliptic Curve ElGamal is homomorphic with respect to the addition as shown in Equation \ref{eq:ecelgamal_summary}.

\begin{equation}
\label{eq:ecelgamal_summary}
    {\varepsilon(m_1, r_1)} + {\varepsilon(m_2, r_2)}
    =
    {\varepsilon(m_1 + m_2, r_1 + r_2)}
\end{equation}

In elliptic curve ElGamal, plaintexts are casted to points on the given curve and the ciphertexts are points as well. Decryption of a ciphertext will give the cast point. Restoring plaintext from that cast point requires to solve elliptic curve discrete logarithm problem which is hard. Still, restoration calculation can be done fast if the plaintext is millionish number.

\subsubsection{Scalar Multiplication Feature}

Finding the k-th multiple of a ciphertext is calculated in Equations \ref{eq:ecelgamal_scalar_1} and \ref{eq:ecelgamal_scalar_2}.

\begin{equation}
\label{eq:ecelgamal_scalar_1}
    k \times {\varepsilon(m_1, r_1)} = k \times (r_1 \times G, r_1 \times Q + m_1 \times G)
\end{equation}

\begin{equation}
\label{eq:ecelgamal_scalar_2}
    k \times {\varepsilon(m_1, r_1)} = (k \times r_1 \times G, k \times r_1 \times Q + k \times m_1 \times G)
\end{equation}

On the other hand, encryption of k-th multiple of a plaintext $m_1$ with a random key k-th multiple of $r_1$ will give the same result as shown in Equation \ref{eq:ecelgamal_scalar_3}. This proves the scalar multiplication property of elliptic curve ElGamal.

\begin{equation}
\label{eq:ecelgamal_scalar_3}
    {\varepsilon(k \times m_1, k \times r_1)} = (k\times r_1 \times G, k \times r_1 \times Q + k\times m_1 \times G)
\end{equation}

After decrypting,  only the first argument, which is the plain message, of the encryption function will be recovered and the random key in the second argument will be entirely removed.


\subsection{Paillier}
\label{sec:paillier}

Paillier cryptosystem was intented by Pascal Paillier in 1999 \cite{paillier}. It is depending on the difficulty of computing n-th residue classes. A general encryption procedure of Paillier algorithm is shown in Equation \ref{eq:paillier_generic} where m is plaintext, r is random key, g is generator and n is RSA modulus.

\begin{equation}
\label{eq:paillier_generic}
    {\varepsilon(m, r)} = ({ g^m \times r^n }) \quad mod \quad n^2
\end{equation}

Then, encryption of plaintext pairs $m_1$ and $m_2$ with random keys $r_1$ and $r_2$ will be calculated using Equations \ref{eq:paillier_c1} and \ref{eq:paillier_c2}.

\begin{equation}
\label{eq:paillier_c1}
    {\varepsilon(m_1, r_1)} = ({ g^{m_1} \times {r_1}^n }) \quad mod \quad n^2
\end{equation}

\begin{equation}
\label{eq:paillier_c2}
    {\varepsilon(m_2, r_2)} = ({ g^{m_2} \times {r_2}^n }) \quad mod \quad n^2
\end{equation}

Thereafter, multiplication of ciphertexts will be calculated using Equation \ref{eq:paillier_c1_times_c2}.

\begin{equation}
\label{eq:paillier_c1_times_c2}
    {\varepsilon(m_1, r_1)} \times {\varepsilon(m_2, r_2)} =  ({ g^{m_1} \times {r_1}^n }) \times ({ g^{m_2} \times {r_2}^n }) \quad mod \quad n^2
\end{equation}

The multiplication can be reorganized as shown in Equations \ref{eq:paillier_c1_times_c2_reorganized} and \ref{eq:paillier_c1_times_c2_reorganized_2}.

\begin{equation}
\label{eq:paillier_c1_times_c2_reorganized}
    {\varepsilon(m_1, r_1)} \times {\varepsilon(m_2, r_2)} =  ({ g^{m_1} \times g^{m_2} \times {r_1}^n \times {r_2}^n }) \quad mod \quad n^2
\end{equation}

\begin{equation}
\label{eq:paillier_c1_times_c2_reorganized_2}
    {\varepsilon(m_1, r_1)} \times {\varepsilon(m_2, r_2)} =  ( g^{m_1 + m_2} \times {(r_1 \times r_2)}^n) \quad mod \quad n^2
\end{equation}

On the other hand, encryption of addition of plaintexts $m_1$ and $m_2$ with random key $r_1$ $\times$ $r_2$ will give same result as shown in Equation \ref{eq:paillier_encryption_of_m1_plus_m2}.

\begin{equation}
\label{eq:paillier_encryption_of_m1_plus_m2}
    {\varepsilon(m_1 + m_2, r_1 \times r_2)} =  ( g^{m_1 + m_2} \times {(r_1 \times r_2)}^n) \quad mod \quad n^2
\end{equation}

In conclusion, Paillier is homomorphic with respect to the addition as shown in Equation \ref{eq:paillier_summary}.

\begin{equation}
\label{eq:paillier_summary}
    {\varepsilon(m_1, r_1)} \times {\varepsilon(m_2, r_2)} = {\varepsilon(m_1 + m_2, r_1 \times r_2)}
\end{equation}

\subsubsection{Ciphertext Regeneration Feature}

Neutral element in addition is 0. So, if the neutral element is encrypted and then homomorphic addition applied to a ciphertext encrypted by Paillier, corresponding plaintext should be same whereas ciphertext changes. In other words, a plaintext may have many ciphertexts. This basic feature allows us to regenerate ciphertexts as shown in Equations \ref{eq:paillier_neutral_1} and \ref{eq:paillier_neutral_2} where \textit{D} denotes decryption.

\begin{equation}
\label{eq:paillier_neutral_1}
    {\varepsilon(m_1, r_1)} \times {\varepsilon(0, r_2)} \neq {\varepsilon(m_1, r_1)}
\end{equation}

\begin{equation}
\label{eq:paillier_neutral_2}
    D({\varepsilon(m_1, r_1)} \times {\varepsilon(0, r_2)}) = D({\varepsilon(m_1, r_1)})
\end{equation}

\subsubsection{Scalar Multiplication Feature}

Even though Paillier cryptosystem is not homomorphic with respect to the multiplication, it allows to multiply a ciphertext with a known constant aka scalar multiplication. This can be calculated with adding a ciphertext into itself a constant k times but this operation's complexity is $O(k)$. On the other hand, calculating a ciphertext to the power of that constant will give same result as calculated in Equations \ref{eq:paillier_scalar_1} and \ref{eq:paillier_scalar_2}.

\begin{equation}
\label{eq:paillier_scalar_1}
    {\varepsilon(m_1, r_1)}^k = ({ g^{m_1} \times {r_1}^n })^k \quad mod \quad n^2
\end{equation}

\begin{equation}
\label{eq:paillier_scalar_2}
    {\varepsilon(m_1, r_1)}^k = ({ g^{m_1 \times k} \times {r_1}^{n \times k} }) \quad mod \quad n^2
\end{equation}

On the other hand, multiplication of $m_1$ and k with the random key k-th power of $r_1$ will give same result as shown in Equation \ref{eq:paillier_scalar_3}.

\begin{equation}
\label{eq:paillier_scalar_3}
    {\varepsilon(m_1 \times k, r_1^k)} = ({ g^{m_1 \times k} \times {r_1}^{n \times k} }) \quad mod \quad n^2
\end{equation}

After decryption, only the first argument of the encryption function will be recovered which is plaintext times constant value.


\subsection{Damgard-Jurik}
\label{sec:damgard}

Damgard-Jurik is a generalized type of Paillier algorithm invented by Ivan Damgard and Mads Jurik in 2001 \cite{damgard}. Pailier uses computations modulo $n^2$ whereas Damgard-Jurik uses $n^{s+1}$ where n is a RSA modulus. In other words, Paillier is a subset of Damgard-Jurik where s equals to 1. Similar to Paillier, it is depending on the difficulty of computing n-th residue classes. A generalized encryption calculation of the cryptosystem is shown in Equation \ref{eq:damgard_generic}.

\begin{equation}
\label{eq:damgard_generic}
    {\varepsilon(m, r)} = ({ g^m \times r^{n^s} }) \quad mod \quad n^{s+1}
\end{equation}

Then, encryption of plaintext pairs $m_1$ and $m_2$ with random keys $r_1$ and $r_2$ will be calculated using Equations \ref{eq:damgard_c1} and \ref{eq:damgard_c2}.

\begin{equation}
\label{eq:damgard_c1}
    {\varepsilon(m_1, r_1)} = ({ g^{m_1} \times {r_1}^{n^s} }) \quad mod \quad n^{s+1}
\end{equation}

\begin{equation}
\label{eq:damgard_c2}
    {\varepsilon(m_2, r_2)} = ({ g^{m_2} \times {r_2}^{n^s} }) \quad mod \quad n^{s+1}
\end{equation}

Thereafter, multiplication of ciphertext will be calculated using Equation.

\begin{equation}
\label{eq:damgard_c1_times_c2}
    {\varepsilon(m_1, r_1)} \times {\varepsilon(m_2, r_2)} =  ({ g^{m_1} \times {r_1}^{n^s} }) \times ({ g^{m_2} \times {r_2}^{n^s} }) \quad mod \quad n^{s+1}
\end{equation}

The multiplication can be reorganized as shown in Equations \ref{eq:damgard_c1_times_c2_reorganized} and \ref{eq:damgard_c1_times_c2_reorganized_2}.

\begin{equation}
\label{eq:damgard_c1_times_c2_reorganized}
    {\varepsilon(m_1, r_1)} \times {\varepsilon(m_2, r_2)} =  ({ g^{m_1} \times g^{m_2} \times {r_1}^{n^s} \times {r_2}^{n^s} }) \quad mod \quad n^{s+1}
\end{equation}

\begin{equation}
\label{eq:damgard_c1_times_c2_reorganized_2}
    {\varepsilon(m_1, r_1)} \times {\varepsilon(m_2, r_2)} =  ( g^{m_1 + m_2} \times {(r_1 \times r_2)}^{n^s}) \quad mod \quad n^{s+1}
\end{equation}

On the other hand, encryption of addition of plaintexts $m_1$ and $m_2$ with random key $r_1$ $\times$ $r_2$ will give same result as shown in Equation \ref{eq:damgard_encryption_of_m1_plus_m2}.

\begin{equation}
\label{eq:damgard_encryption_of_m1_plus_m2}
    {\varepsilon(m_1 + m_2, r_1 \times r_2)} =  ( g^{m_1 + m_2} \times {(r_1 \times r_2)}^{n^s}) \quad mod \quad n^{s+1}
\end{equation}

In conclusion, Damgard-Jurik is homomorphic with respect to the addition as shown in Equation \ref{eq:damgard_summary}.

\begin{equation}
\label{eq:damgard_summary}
    {\varepsilon(m_1, r_1)} \times {\varepsilon(m_2, r_2)} = {\varepsilon(m_1 + m_2, r_1 \times r_2)}
\end{equation}

\subsubsection{Ciphertext Regeneration Feature}

Encrypting the neutral element of addition which is 0 and then adding it to a ciphertext will regenerate ciphertext whereas corresponding plaintext remains same as shown in Equation \ref{eq:damgard_neutral_1} and \ref{eq:damgard_neutral_2} where \textit{D} denotes decryption.

\begin{equation}
\label{eq:damgard_neutral_1}
    {\varepsilon(m_1, r_1)} \times {\varepsilon(0, r_2)} \neq {\varepsilon(m_1, r_1)}
\end{equation}

\begin{equation}
\label{eq:damgard_neutral_2}
    D({\varepsilon(m_1, r_1)} \times {\varepsilon(0, r_2)}) = D({\varepsilon(m_1, r_1)})
\end{equation}

\subsubsection{Scalar Multiplication Feature}

Even though Damgard-Jurik cryptosystem is not multiplicatively homomorphic, it has scalar multiplication feature. Finding the k-th power of a ciphertext where k is a constant will be calculated as shown Equations \ref{eq:damgard_scalar_1} and \ref{eq:damgard_scalar_2}.

\begin{equation}
\label{eq:damgard_scalar_1}
    {\varepsilon(m_1, r_1)}^k = ({ g^{m_1} \times {r_1}^(n^s) })^k \quad mod \quad n^{s+1}
\end{equation}

\begin{equation}
\label{eq:damgard_scalar_2}
    {\varepsilon(m_1, r_1)}^k = ({ g^{m_1 \times k} \times {r_1}^{(n^s \times k)} }) \quad mod \quad n^{s+1}
\end{equation}

On the other hand, encryption of m1 times k with random key r1 to the power of k will give same calculation as shown in \ref{eq:damgard_scalar_3}.

\begin{equation}
\label{eq:damgard_scalar_3}
    {\varepsilon(m_1 \times k, r_1^k)} = ({ g^{m_1 \times k} \times {r_1}^{(k \times n^s)} }) \quad mod \quad n^{s+1}
\end{equation}

The decryption process will restore just the first argument. The second argument, which contains the random number, will be completely discarded.


\subsection{Okamoto-Uchiyama}
\label{sec:okamoto}

Okamoto-Uchiyama was invented by Tatsuaki Okamoto and Shigenori Uchiyama in 1998 \cite{okamoto}. General encryption rule of Okamoto-Uchiyama cryptosystem is shown in Equation \ref{eq:okamoto_generic}.

\begin{equation}
\label{eq:okamoto_generic}
    {\varepsilon(m, r)} = ({ g^m \times h^r }) \quad mod \quad n
\end{equation}

Then, encryption of plaintext pairs $m_1$ and $m_2$ with random keys $r_1$ and $r_2$ will be calculated using Equations \ref{eq:okamoto_c1} and \ref{eq:okamoto_c2}.

\begin{equation}
\label{eq:okamoto_c1}
    {\varepsilon(m_1, r_1)} = ({ g^{m_1} \times {h}^{r_1} }) \quad mod \quad n
\end{equation}

\begin{equation}
\label{eq:okamoto_c2}
    {\varepsilon(m_2, r_2)} = ({ g^{m_2} \times {h}^{r_2} }) \quad mod \quad n
\end{equation}

Thereafter, multiplication of ciphertext will be calculated using Equation \ref{eq:okamoto_c1_times_c2}.

\begin{equation}
\label{eq:okamoto_c1_times_c2}
    {\varepsilon(m_1, r_1)} \times {\varepsilon(m_2, r_2)} 
    =
    ({ g^{m_1} \times {h}^{r_1} }) \times ({ g^{m_2} \times {h}^{r_2} }) \quad mod \quad n
\end{equation}

The multiplication can be reorganized as shown in Equations \ref{eq:okamoto_c1_times_c2_reorganized} and \ref{eq:okamoto_c1_times_c2_reorganized_2}.

\begin{equation}
\label{eq:okamoto_c1_times_c2_reorganized}
    {\varepsilon(m_1, r_1)} \times {\varepsilon(m_2, r_2)} 
    =
    ({ g^{m_1} \times g^{m_2} \times {h}^{r_1} \times {h}^{r_2} }) \quad mod \quad n
\end{equation}

\begin{equation}
\label{eq:okamoto_c1_times_c2_reorganized_2}
    {\varepsilon(m_1, r_1)} \times {\varepsilon(m_2, r_2)} 
    =
    ({ g^{m_1 + m_2} \times {h}^{r_1 + r_2} }) \quad mod \quad n
\end{equation}

On the other hand, encryption of addition of plaintexts $m_1$ and $m_2$ with random key $r_1$+$r_2$ will give same result as shown in Equation \ref{eq:okamoto_m1_plus_m2_encrypted}.

\begin{equation}
\label{eq:okamoto_m1_plus_m2_encrypted}
    {\varepsilon(m_1 + m_2, r_1 + r_2)} 
    =
    ({ g^{m_1 + m_2} \times {h}^{r_1 + r_2} }) \quad mod \quad n
\end{equation}

In conclusion, Okamoto-Uchiyama is homomorphic with respect to the addition as shown in Equation \ref{eq:okamoto_summary}.

\begin{equation}
\label{eq:okamoto_summary}
    {\varepsilon(m_1, r_1)} \times {\varepsilon(m_2, r_2)}
    =
    {\varepsilon(m_1 + m_2, r_1 + r_2)}
\end{equation}

\subsubsection{Ciphertext Regeneration Feature}

Encrypting the neutral element of addition which is 0 and then adding it to a ciphertext will regenerate ciphertext whereas corresponding plaintext remains same as shown in Equations \ref{eq:okamoto_neutral_1} and \ref{eq:okamoto_neutral_2} where \textit{D} denotes decryption.

\begin{equation}
\label{eq:okamoto_neutral_1}
    {\varepsilon(m_1, r_1)} \times {\varepsilon(0, r_2)}
    \neq
    {\varepsilon(m_1, r_1)}
\end{equation}

\begin{equation}
\label{eq:okamoto_neutral_2}
    D({\varepsilon(m_1, r_1)} \times {\varepsilon(0, r_2)})
    =
    D({\varepsilon(m_1, r_1)})
\end{equation}

\subsubsection{Scalar Multiplication Feature}

Okamoto-Uchiyama supports scalar multiplication as shown in Equations \ref{eq:okamoto_scalar_1} and \ref{eq:okamoto_scalar_2}.

\begin{equation}
\label{eq:okamoto_scalar_1}
    {\varepsilon(m_1, r_1)}^k = ({ g^{m_1} \times {h}^{r_1} })^k \quad mod \quad n
\end{equation}

\begin{equation}
\label{eq:okamoto_scalar_2}
    {\varepsilon(m_1, r_1)}^k = ({ g^{m_1 \times k} \times {h}^{r_1 \times k} }) \quad mod \quad n
\end{equation}

On the other hand, encryption of $m_1$ times k with random key $r_1$ to the power of k will give same calculation as shown in \ref{eq:damgard_scalar_3}.

\begin{equation}
\label{eq:okamoto_scalar_3}
    {\varepsilon(m_1 \times k, r_1 \times k)} = ({ g^{m_1 \times k} \times {h}^{r_1 \times k} }) \quad mod \quad n
\end{equation}

Once the data is decrypted, the random number in the second argument is ignored.


\subsection{Goldwasser–Micali}
\label{sec:goldwasser}

Goldwasser-Micali was invented by Shafi Goldwasser and Silvio Micali in 1982 \cite{golwasser}. The cryptosystem shows homomorphic features with respect to the exclusive or (XOR). A generalized encryption formula of Goldwasser-Micali is shown in Equation \ref{eq:goldwasser_generic} where b is the bit value, r is a random integer and x is a non-residue number. Unsimilar to other cryptosystems, Goldwasser-Micali encrypts bits instead of plaintext numbers.

\begin{equation}
\label{eq:goldwasser_generic}
    {\varepsilon(b, r)} = ({ r^2 \times x^b }) \quad mod \quad n
\end{equation}

Then, encryption of plain bit text pairs $b_1$ and $b_2$ with random keys $r_1$ and $r_2$ will be calculated using Equations \ref{eq:golwasser_c1} and \ref{eq:golwasser_c2}.

\begin{equation}
\label{eq:golwasser_c1}
    {\varepsilon(b_1, r_1)} = ({ {r_1}^2 \times x^{b_1} }) \quad mod \quad n
\end{equation}

\begin{equation}
\label{eq:golwasser_c2}
    {\varepsilon(b_2, r_2)} = ({ {r_2}^2 \times x^{b_2} }) \quad mod \quad n
\end{equation}

Thereafter, multiplication of ciphertexts will be calculated using Equation \ref{eq:golwasser_c1_times_c2}.

\begin{equation}
\label{eq:golwasser_c1_times_c2}
    {\varepsilon(b_1, r_1)} \times {\varepsilon(b_2, r_2)} = ({ {r_1}^2 \times x^{b_1} }) \times ({ {r_2}^2 \times x^{b_2} }) \quad mod \quad n
\end{equation}

The multiplication can be reorganized as shown in Equations \ref{eq:golwasser_c1_times_c2_reorganized} and \ref{eq:golwasser_c1_times_c2_reorganized_2}.

\begin{equation}
\label{eq:golwasser_c1_times_c2_reorganized}
    {\varepsilon(b_1, r_1)} \times {\varepsilon(b_2, r_2)} = ({ {r_1}^2 \times {r_2}^2 \times x^{b_1} \times x^{b_2} }) \quad mod \quad n
\end{equation}

\begin{equation}
\label{eq:golwasser_c1_times_c2_reorganized_2}
    {\varepsilon(b_1, r_1)} \times {\varepsilon(b_2, r_2)} = { {(r_1 \times r_2)}^2 \times x^{b_1 + b_2} } \quad mod \quad n
\end{equation}

On the other hand, encryption of $b_1$ + $b_2$ with random key $r_1$ $\times$ $r_2$ will give same result as shown in Equation \ref{eq:golwasser_m1_plus_m2_encrypted}. Herein, addition on bits is equivalent to xor operation.

\begin{equation}
\label{eq:golwasser_m1_plus_m2_encrypted}
    {\varepsilon(b_1 \oplus b_2, r_1 \times r_2)} = { ({r_1 \times r_2})^2 \times x^{b_1 + b_2} } \quad mod \quad n
\end{equation}

In conclusion, Goldwasser-Micali is homomorphic with respect to the exclusive or (XOR) as shown in Equation \ref{eq:goldwasser_summary}.

\begin{equation}
\label{eq:goldwasser_summary}
    {\varepsilon(b_1, r_1)} \times {\varepsilon(b_2, r_2)} = {\varepsilon(b_1 \oplus b_2, r_1 \times r_2)}
\end{equation}


\subsection{Benaloh}
\label{sec:benaloh}

Benaloh cryptosystem was invented by Josh Benaloh in 1985 \cite{benaloh}. It is an extension of Goldwasser-Micali \cite{golwasser} cryptosystem. Benaloh shows additively homomorphic features whereas Goldwasser-Micali was homomorphic with respect to the exclusive or (XOR). A generalized encryption formula of Benaloh is shown in Equation \ref{eq:benaloh_generic}.

\begin{equation}
\label{eq:benaloh_generic}
    {\varepsilon(m, r)} = ({ y^m \times u^r }) \quad mod \quad n
\end{equation}

Then, encryption of plaintext pairs $m_1$ and $m_2$ with random keys $r_1$ and $r_2$ will be calculated using Equations \ref{eq:benaloh_c1} and \ref{eq:benaloh_c2}.

\begin{equation}
\label{eq:benaloh_c1}
    {\varepsilon(m_1, r_1)} = ({ y^{m_1} \times u^{r_1} }) \quad mod \quad n
\end{equation}

\begin{equation}
\label{eq:benaloh_c2}
    {\varepsilon(m_2, r_2)} = ({ y^{m_2} \times u^{r_2} }) \quad mod \quad n
\end{equation}

Thereafter, multiplication of ciphertexts will be calculated using Equation \ref{eq:benaloh_c1_times_c2}.

\begin{equation}
\label{eq:benaloh_c1_times_c2}
    {\varepsilon(m_1, r_1)} \times {\varepsilon(m_2, r_2)} = ({ y^{m_1} \times u^{r_1} }) \times ({ y^{m_2} \times u^{r_2} }) \quad mod \quad n
\end{equation}

The multiplication can be reorganized as shown in Equations \ref{eq:benaloh_c1_times_c2_reorganized} and \ref{eq:benaloh_c1_times_c2_reorganized_2}.

\begin{equation}
\label{eq:benaloh_c1_times_c2_reorganized}
    {\varepsilon(m_1, r_1)} \times {\varepsilon(m_2, r_2)} = ({ y^{m_1} \times y^{m_2} \times u^{r_1} \times u^{r_2} }) \quad mod \quad n
\end{equation}

\begin{equation}
\label{eq:benaloh_c1_times_c2_reorganized_2}
    {\varepsilon(m_1, r_1)} \times {\varepsilon(m_2, r_2)} = ({ y^{m_1 + m_2} \times u^{r_1 + r_2}}) \quad mod \quad n
\end{equation}

On the other hand, encryption of addition of plaintexts $m_1$ and $m_2$ with random key $r_1$+$r_2$ will give same result as shown in Equation \ref{eq:benaloh_m1_plus_m2_encrypted}.

\begin{equation}
\label{eq:benaloh_m1_plus_m2_encrypted}
    {\varepsilon(m_1 + m_2, r_1 + r_2)} = ({ y^{m_1 + m_2} \times u^{r_1 + r_2}}) \quad mod \quad n
\end{equation}

In conclusion, Benaloh is homomorphic with respect to the addition as shown in Equation \ref{eq:benaloh_summary}.

\begin{equation}
\label{eq:benaloh_summary}
    {\varepsilon(m_1, r_1)} \times {\varepsilon(m_2, r_2)}
    =
    {\varepsilon(m_1 + m_2, r_1 + r_2)}
\end{equation}

Similar to exponential ElGamal and elliptic curve ElGamal, decryption of Benaloh is difficult problem. It requires to solve discrete logarithm to restore the plaintext.

\subsubsection{Ciphertext Regeneration Feature}

Encrypting the neutral element of addition which is 0 and then adding it to a ciphertext will regenerate ciphertext whereas corresponding plaintext remains same as shown in Equations \ref{eq:benaloh_neutral_1} and \ref{eq:benaloh_neutral_2} where \textit{D} denotes decryption.

\begin{equation}
\label{eq:benaloh_neutral_1}
    {\varepsilon(m_1, r_1)} \times {\varepsilon(0, r_2)}
    \neq
    {\varepsilon(m_1, r_1)}
\end{equation}

\begin{equation}
\label{eq:benaloh_neutral_2}
    D({\varepsilon(m_1, r_1)} \times {\varepsilon(0, r_2)})
    =
    D({\varepsilon(m_1, r_1)})
\end{equation}

\subsubsection{Scalar Multiplication Feature}

Finding the k-th power of ciphertext will be equal to the scalar multiplication of k and plaintext message as shown in Equations \ref{eq:benaloh_scalar_1} and \ref{eq:benaloh_scalar_2}.

\begin{equation}
\label{eq:benaloh_scalar_1}
    {\varepsilon(m_1, r_1)}^k = ({ y^{m_1} \times u^{r_1} })^k \quad mod \quad n
\end{equation}

\begin{equation}
\label{eq:benaloh_scalar_2}
    {\varepsilon(m_1, r_1)}^k = ({ y^{m_1 \times k} \times u^{r_1 \times k} }) \quad mod \quad n
\end{equation}

On the other hand, encryption of k times m1 with random key r1 times k will be same as shown in Equation \ref{eq:benaloh_scalar_3}.

\begin{equation}
\label{eq:benaloh_scalar_3}
    {\varepsilon(m_1 \times k, r_1 \times k)} = ({ y^{m_1 \times k} \times u^{r_1 \times k} }) \quad mod \quad n
\end{equation}

Once decryption is done, just the first argument of encryption function will be restored,and the random number in the second argument will be dropped.


\subsection{Naccache-Stern}
\label{sec:naccache}

Naccache-Stern was invented by David Naccache and Jacques Stern in 1998 \cite{naccache}. It is depending on the difficulty of higher residuosity problem. A generalized encryption procedure of the cryptosystem is shown in Equation \ref{eq:naccache_generic}.

\begin{equation}
\label{eq:naccache_generic}
    {\varepsilon(m, \sigma)} = ({ g^m \times r^\sigma }) \quad mod \quad n
\end{equation}

Then, encryption of plaintext pairs $m_1$ and $m_2$ with random keys $r_1$ and $r_2$ will be calculated using Equations \ref{eq:naccache_c1} and \ref{eq:naccache_c2}.

\begin{equation}
\label{eq:naccache_c1}
    {\varepsilon(m_1, \sigma_1)} = ({ g^{m_1} \times r^{\sigma_1} }) \quad mod \quad n
\end{equation}

\begin{equation}
\label{eq:naccache_c2}
    {\varepsilon(m_2, \sigma_2)} = ({ g^{m_2} \times r^{\sigma_2} }) \quad mod \quad n
\end{equation}

Thereafter, multiplication of ciphertexts will be calculated using Equation \ref{eq:naccache_c1_times_c2}.

\begin{equation}
\label{eq:naccache_c1_times_c2}
    {\varepsilon(m_1, \sigma_1)} \times {\varepsilon(m_2, \sigma_2)} =  ({ g^{m_1} \times r^{\sigma_1} }) \times ({ g^{m_2} \times r^{\sigma_2} }) \quad mod \quad n
\end{equation}

The multiplication can be reorganized as shown in Equation \ref{eq:naccache_c1_times_c2_reorganized}.

\begin{equation}
\label{eq:naccache_c1_times_c2_reorganized}
    {\varepsilon(m_1, \sigma_1)} \times {\varepsilon(m_2, \sigma_2)} =  ({ g^{m_1 + m_2} \times r^{\sigma_1 + \sigma_2} }) \quad mod \quad n
\end{equation}

On the other hand, encryption of addition of plaintexts $m_1$ and $m_2$ will give same result as shown in Equation \ref{eq:naccache_m1_plus_m2_encrypted}.

\begin{equation}
\label{eq:naccache_m1_plus_m2_encrypted}
    {\varepsilon(m_1 + m_2, \sigma_1 + \sigma_2)} =  ({ g^{m_1 + m_2} \times r^{\sigma_1 + \sigma_2} }) \quad mod \quad n
\end{equation}

In conclusion, Naccache-Stern is homomorphic with respect to the addition as shown in Equation \ref{eq:naccache_summary}.

\begin{equation}
\label{eq:naccache_summary}
    {\varepsilon(m_1, \sigma_1)} \times {\varepsilon(m_2, \sigma_2)} = {\varepsilon(m_1 + m_2, \sigma_1 + \sigma_2 )} 
\end{equation}

Similar to Benaloh, decryption of Naccahe-Stern requires to solve discete logarithm problem.

\subsubsection{Ciphertext Regeneration Feature}

Encrypting the neutral element of addition which is 0 and then adding it to a ciphertext will regenerate ciphertext whereas corresponding plaintext remains same as shown in Equations \ref{eq:naccache_neutral_1} and \ref{eq:naccache_neutral_2} where \textit{D} denotes decryption.

\begin{equation}
\label{eq:naccache_neutral_1}
    {\varepsilon(m_1, \sigma_1)} \times {\varepsilon(0, \sigma_2)} \neq {\varepsilon(m_1, \sigma_1)}
\end{equation}

\begin{equation}
\label{eq:naccache_neutral_2}
    D({\varepsilon(m_1, \sigma_1)} \times {\varepsilon(0, \sigma_2)}) = D({\varepsilon(m_1, \sigma_1)})
\end{equation}

\subsubsection{Scalar Multiplication Feature}

Naccache-Stern cryptosystem shows scalar multiplicationf features. Find the k-th power of a ciphertext will be equal the encryption of k times plaintext as shown in Equations \ref{eq:naccache_scalar_1}, \ref{eq:naccache_scalar_2} and \ref{eq:naccache_scalar_3}.

\begin{equation}
\label{eq:naccache_scalar_1}
    {\varepsilon(m_1, \sigma)}^k = ({ g^{m_1} \times r^{\sigma} })^k \quad mod \quad n
\end{equation}

\begin{equation}
\label{eq:naccache_scalar_2}
    {\varepsilon(m_1, \sigma)}^k = ({ g^{m_1 \times k} \times r^{\sigma \times k} }) \quad mod \quad n
\end{equation}

\begin{equation}
\label{eq:naccache_scalar_3}
    {\varepsilon(m_1 \times k, \sigma \times k)} = ({ g^{m_1 \times k} \times r^{\sigma \times k} }) \quad mod \quad n
\end{equation}

During decryption, the first argument is retrieved while the second argument is omitted.

\section{Design and Architecture}
\label{sec:software_architecture}

An abstract class in programming is a class that cannot be instantiated on its own and is designed to be subclassed, typically containing one or more abstract methods that must be implemented by its subclasses \cite{abstract}. The LightPHE framework is designed with a generic abstract class called Homomorphic. This class will include all the necessary methods for homomorphic operations, such as the addition of ciphertexts, multiplication of ciphertexts, xor of ciphertext, scalar multiplication of a ciphertext with a constant, and ciphertext regeneration. In addition to these homomorphic operations, the Homomorphic abstract class will have methods for generating key pairs for the cryptosystem, as well as methods for encryption and decryption.

Besides, each PHE algorithm generates different type of ciphertexts. Basically, this specifies the design of the class. Table \ref{tab:classes} shows ciphertext types of these algorithms and also the random key availability in the encryption process.

\begin{minipage}{\linewidth} 
\begin{table}[H]
\caption{Ciphertext Types and Random Key Requirements of PHE Algorithms}
\centering
\begin{tabular}{lcc}
\toprule
Algorithm & Ciphertext Type & \begin{tabular}[c]{@{}c@{}}Encryption Requires\\ Random Key\end{tabular} \\
\midrule
RSA                    & int                     & No \\
Paillier               & int                     & Yes \\
DamgardJurik           & int                     & Yes \\
OkamotoUchiyama        & int                     & Yes \\
Benaloh                & int                     & Yes \\
NaccacheStern          & int                     & Yes \\
GoldwasserMicali       & List{[}int{]}           & Yes \\
ElGamal                & Tuple{[}int, int{]}     & Yes \\
Exp. ElGamal           & Tuple{[}int, int{]}     & Yes \\
EC ElGamal & Tuple{[}Tuple{[int, int]}, Tuple{[int, int]}{]} & Yes \\
\bottomrule
\end{tabular}
\label{tab:classes}
\end{table}
\end{minipage}
\newline
\newline

\begin{sidewaysfigure}[ht]
    \includegraphics[width=\textwidth]{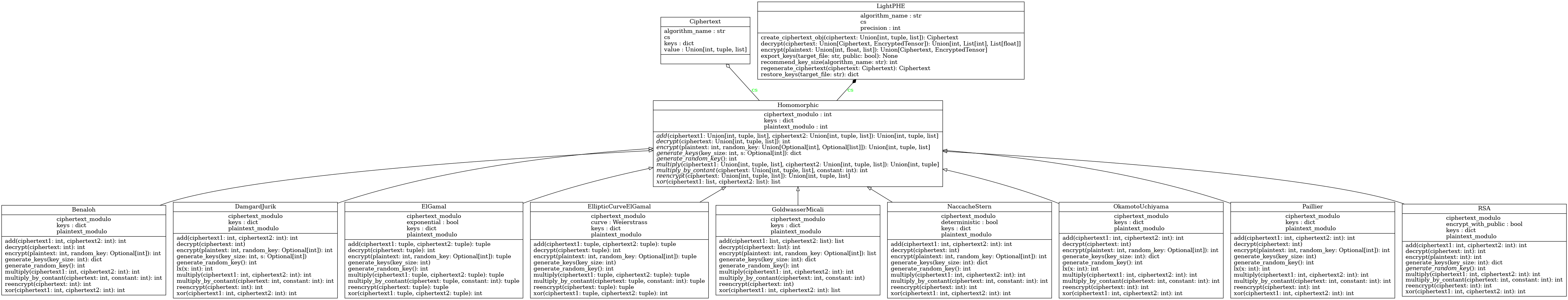}
    \caption{UML Class Diagram of LightPHE}
    \label{fig:cs_classes}
\end{sidewaysfigure}

Each partially homomorphic encryption (PHE) algorithm will have its own class that inherits from the Homomorphic abstract class as shown in Figure \ref{fig:cs_classes}. Abstract class covers all ciphertext types whereas PHE algorithms' classes followed the design mentioned in Table \ref{tab:classes}.

For additively homomorphic algorithms, the framework will throw exceptions in PHE classes for homomorphic multiplication and homomorphic XOR operations, as these are not supported by these algorithms. For multiplicatively homomorphic algorithms, the framework will throw exceptions for homomorphic addition and homomorphic XOR operations. Additionally, exceptions will be thrown for scalar multiplication and ciphertext regeneration because these operations are only supported by additively homomorphic algorithms. For exclusively homomorphic algorithms, the framework will throw exceptions for both homomorphic addition and homomorphic multiplication. Similarly, exceptions will be thrown for scalar multiplication and ciphertext regeneration because these operations are only supported by additively homomorphic algorithms. In that way, each PHE algorithm class will have same methods.

Users will only interact with LightPHE class and they do not have to play with backend cryptosystems' classes at all. Once a cryptosystem created, it will have its own encrypt and decrypt methods. Encrypt method will return a Ciphertext class. Basically, the returned values from backend cryptosystems are set to the value argument of the Ciphertext class. We overwritten the addition, multiplication and xor method on this class. In that way, adding or multiplying two Ciphertexts classes will perform homomorphic addition, multiplication or xor. Similarly, right multiplication method is overwritten to perform scalar multiplication when multiplying a Ciphertext object with a constant.

\section{Implementation}
\label{sec:implementation}

To use the LightPHE framework, the library is imported and the LightPHE class is initialized with the desired partially homomorphic encryption (PHE) algorithm. The following code example demonstrates this process.

\begin{minipage}{\linewidth} 
\begin{lstlisting}[frame=tb, caption=Building a Cryptosystem, label=building_cs, language=Python]
# install the library if not installed yet
!pip install lightphe

# import the library
from lightphe import LightPHE

# supported PHE algorithms
algorithms = [
    "RSA",
    "ElGamal",
    "Goldwasser-Micali",
    "Exponential-ElGamal",
    "EllipticCurve-ElGamal",
    "Paillier",
    "Damgard-Jurik",
    "Okamoto-Uchiyama",
    "Benaloh",
    "Naccache-Stern",
]

# build cryptosystem with private public key pair
cs = LightPHE(
    algorithm_name = algorithms[0],
    key_size = 1024,
)
\end{lstlisting}
\end{minipage}
\newline
\newline

In the example shown in Algorithm \ref{building_cs}, the LightPHE class is imported from the lightphe library. The algorithms list contains the names of all the supported PHE algorithms. Initializing the LightPHE class with first algorithm in the list, which corresponds to RSA in this case, generates a random private-public key pair for the chosen cryptosystem. This procedure is sufficient to build a cryptosystem using the selected PHE algorithm.

Then, the following code demonstrates defining a plaintext value, encrypting it, and then decrypting it to verify correctness.

\begin{minipage}{\linewidth} 
\begin{lstlisting}[frame=tb, caption=Encrypt and Decrypt with Built Cryptosystem, label=encrypt_and_decrypt, language=Python]
# define plaintext
m = 17

# calculate ciphertext with public key
c = cs.encrypt(m)

# proof of work - decrypt with private key
assert cs.decrypt(c) == m
\end{lstlisting}
\end{minipage}
\newline
\newline

In the example shown in Algorithm \ref{encrypt_and_decrypt}, A plaintext value m is defined and set to 17. The encrypt method of the cryptosystem is called with plaintext, producing the ciphertext c. The decrypt method of cryptosystem is then called with ciphertext as the argument to decrypt the ciphertext. An assertion is made to verify that the decrypted value matches the original plaintext m. This ensures that the encryption and decryption processes are functioning correctly within the cryptosystem.

The following code demonstrates homomorphic addition using the Paillier algorithm:

\begin{minipage}{\linewidth} 
\begin{lstlisting}[frame=tb, caption=Homomorphic Addition, label=homomorphic_addition, language=Python]
# build Paillier cryptosystem - it's additively homomorphic
cs = LightPHE(algorithm_name = "Paillier")

# define plaintexts
m1 = 10000 # base salary
m2 = 500 # wage increase in amount

# calculate ciphertexts - this is done with just public key
c1 = cs.encrypt(m1)
c2 = cs.encrypt(m2)

# homomorphic addition - private key is not required!
c3 = c1 + c2

# proof of work - decrypt with private key
assert cs.decrypt(c3) == m1 + m2
\end{lstlisting}
\end{minipage}
\newline
\newline

In the example demonstrated in \ref{homomorphic_addition}, the LightPHE class is initialized with the Paillier algorithm. Two plaintext values, $m_1$ and $m_2$, are defined and set to 10000, representing the base salary, and 500, representing the wage increase in amount, respectively. The encrypt method of the LightPHE instance cs is called with $m_1$ and $m_2$ as arguments, producing the ciphertexts $c_1$ and $c_2$. Encryption can be done with just public keys, without need of private keys. 

Types of $c_1$ and $c_2$ are Ciphertext and LightPHE overwritten that class' addition method to perform homomorphic addition as mentioned in Section \ref{sec:software_architecture}. So, homomorphic addition is performed by adding $c_1$ and $c_2$ to produce a new ciphertext $c_3$, which represents the updated encrypted salary. This operation does not require the private key, too. 

Finally, an assertion is made to verify that decrypting $c_3$ returns the sum of the original plaintext values ($m_1$ + $m_2$). Decryption can only be done by the data owner who holds private key. This demonstrates the correctness of the homomorphic addition operation.

\begin{minipage}{\linewidth} 
\begin{lstlisting}[frame=tb, caption=Scalar Multiplication, label=scalar_multiplication, language=Python]
# build Paillier cryptosystem - it's additively homomorphic
cs = LightPHE(algorithm_name = "Paillier")

# define base salary
m1 = 10000

# find encrypted base salary
c1 = cs.encrypt(m1)

# set wage increase percentage as constant
k = 1.05

# calculating encrypted updated salary - private key is not required!
c4 = k * c1

# proof of work - decrypt with private key
assert cs.decrypt(c4) == k * m1
\end{lstlisting}
\end{minipage}
\newline
\newline

In the example illustrated in Algorithm \ref{scalar_multiplication}, a scalar k is defined to represent a 5\% wage increase and is set to 1.05. The variable $c_1$ is a type of Ciphertext, representing the encrypted base salary of someone, and the class's right multiplication method is overridden to perform scalar multiplication. Thus, scalar multiplication is performed by multiplying k with the ciphertext $c_1$ to produce a new ciphertext $c_4$, which represents the encrypted updated salary. This operation does not require the private key and can be done by anyone who has the public keys.

Finally, an assertion is made to verify that decrypting $c_4$ returns the product of the scalar k and the original plaintext $m_1$, demonstrating the correctness of the scalar multiplication operation. This operation can only be done by the data owner who holds private key of the cryptosystem.

The following code in Algorithm \ref{unsupported_operation} demonstrates the limitations of the Paillier algorithm with respect to multiplicative and exclusive homomorphic operations:

\begin{minipage}{\linewidth} 
\begin{lstlisting}[frame=tb, caption=Unsupported Operations, label=unsupported_operation, language=Python]
# pailier is not multiplicatively homomorphic
with pytest.raises(ValueError):
  c3 = c1 * c2

# pailier is not exclusively homomorphic
with pytest.raises(ValueError):
  c4 = c1 ^ c2
\end{lstlisting}
\end{minipage}
\newline
\newline

Since the Paillier algorithm is not multiplicatively homomorphic, attempting to multiply ciphertexts $c_1$ and $c_2$ will raise an error with the message "Paillier is not homomorphic with respect to the multiplication". Similarly, since the Paillier algorithm is not exclusively homomorphic, attempting to perform an exclusive OR operation (XOR) on $c_1$ and $c_2$ will raise another error with the message "Paillier is not homomorphic with respect to the exclusive or". These are thrown by the Paillier class' homomorphic multiply and homomorphic xor methods. These exceptions demonstrate the algorithm's limitations concerning these specific homomorphic operations.

\subsection{Performance}
\label{sec:onprem_performance}

The provided Tables \ref{tab:performance_80}, \ref{tab:performance_112}, \ref{tab:performance_128}, \ref{tab:performance_192} showcase the performance metrics of various cryptographic algorithms with default configurations, specifically focusing on key generation, encryption, decryption, and homomorphic operations such as addition, multiplication or xor with different key sizes. These measurements are crucial in evaluating the practicality and efficiency of these algorithms in real-world applications. Here, the encryption and homomorphic operation times are represented in scientific notation to illustrate the performance differences clearly. The values in cells represent the times in seconds and are the average of five different experiments.

In these experiments, 18-bit plaintexts were used to simulate realistic annual salary values, typically ranging in the hundreds of thousands. This choice reflects a practical application scenario where the encryption of such values is relevant.

The tables also adhere to the National Institute of Standards and Technology (NIST) recommendations for key sizes to achieve specific security levels \cite{ecc_keysize} as detailed in Table \ref{tab:nist}. For an 80-bit symmetric key security level, NIST suggests using 160-bit keys for elliptic curve cryptography (ECC) and 1024-bit keys for other algorithms. For a 112-bit symmetric key security level, 224-bit ECC keys and 2048-bit keys for other algorithms are recommended. For a 128-bit symmetric key security level, NIST recommends 256-bit ECC keys and 3072-bit keys for other algorithms. These guidelines ensure that the cryptographic strength meets the required security standards.

\begin{minipage}{\linewidth} 
\begin{table}[H]
\centering
\caption{NIST's Key Size Recommendations}
\begin{tabular}{llll}
\toprule
Symmetric Key Size & RSA Variants Key Size Equivalent & ECC Key Size Equivalent & Expected Lifetime \\
\midrule
80 & 1024 & 160 & Until 2010 \\
112 & 2048 & 224 & Until 2030 \\
128 & 3072 & 256 & Beyond 2030 \\
192 & 7680 & 384 & Much Beyond 2030 \\
\bottomrule
\end{tabular}
\label{tab:nist}
\end{table}
\end{minipage}
\newline
\newline

One notable observation from the tables is the significantly higher decryption times for Elliptic Curve ElGamal and Exponential ElGamal algorithms in particular smaller key sizes. This is attributed to the necessity of solving the Discrete Logarithm Problem (DLP) and Elliptic Curve Discrete Logarithm Problem (ECDLP) during the decryption process. This computational complexity renders these cryptosystems more theoretical than practical, especially given the small plaintext values used in the experiments. As the plaintext size increases, the decryption times are expected to rise even further, exacerbating the impracticality.

Interestingly, larger key sizes do not have a significant impact on the encryption and decryption times of Elliptic Curve ElGamal. This consistency across different key sizes highlights a unique aspect of the algorithm's performance. However, it is important to note that the decryption time for Elliptic Curve ElGamal is still slow when compared to other cryptosystems, making it less practical with today's computational power for applications requiring frequent decryption at the 80, 112, and 128-bit symmetric key size equivalents. On the other hand, Elliptic Curve ElGamal becomes advantageous at the 192-bit symmetric key size equivalent, as it becomes faster in key generation, encryption, and decryption when compared to other additively homomorphic encryption algorithms such as Paillier or Damgard-Jurik. Additionally, Elliptic Curve ElGamal has potential in the post-quantum era because its equivalents have a much greater impact on key generation, encryption, and decryption. While the times for these operations increase exponentially with key size for other algorithms, the times for EC ElGamal increase linearly.

Additionally, it is important to mention that the Benaloh and Naccache-Stern cryptosystems were excluded from these experiments. Despite functioning correctly for smaller key sizes, these algorithms encountered significant issues during the key generation step for 1024-bit and 2048-bit keys, effectively hanging and becoming non-operational. This limitation prevented their inclusion in the comparative analysis, highlighting the challenges in scaling certain cryptosystems to higher security levels.

Homomorphic operations and encryptions, as noted in the tables, are very fast. This efficiency is crucial as these operations are typically handled frequently, with encryption being performed on-premise and homomorphic operations executed in the cloud. In cloud environments, where performance is a key concern, the swift execution of these tasks ensures optimal system performance. Although decryption is slower than both homomorphic operations and encryption, its less frequent occurrence mitigates the impact of its slower speed. It's important to note that key generation occurs only once, and while it may be slightly slower, its infrequent occurrence makes this delay negligible in the overall scheme of cryptographic processes.

Overall, these tables provide valuable insights into the performance and practicality of various cryptographic algorithms under different security settings, emphasizing the balance between theoretical capabilities and real-world applicability.

Finally, all experiments were conducted on a machine with an 11th Gen Intel(R) Core(TM) i7-11370H CPU running at 3.30 GHz with 8 cores. This CPU model belongs to Intel's Tiger Lake family and is commonly found in high-performance laptops. Therefore, the information provided about the CPU ensures transparency regarding the computational environment in which the experiments were conducted.

\begin{minipage}{\linewidth} 
\begin{table}[H]
\caption{Time Consumption of Algorithms with 80-bit Symmetric Key Size Equivalent in Seconds}
\centering
\begin{tabular}{llllll}
\toprule
Algorithm & Key Size & Key Generation & Encrypt & Decrypt & Homomorphic Operation \\ 
\midrule
RSA & 1024 & 0.1483 & \(0.00305 \times 10^{-3}\) & 0.0038 & \(1.70708 \times 10^{-5}\) \\ 
ElGamal & 1024 & 0.0177 & \(1.31831 \times 10^{-3}\) & 0.0007 & \(1.66416 \times 10^{-5}\) \\ 
Paillier & 1024 & 0.0507 & \(1.16231 \times 10^{-2}\) & 0.0120 & \(1.90258 \times 10^{-5}\) \\ 
Damgard-Jurik & 1024 & 0.0585 & \(2.35637 \times 10^{-2}\) & 0.0245 & \(3.82900 \times 10^{-5}\) \\ 
Okamoto-Uchiyama & 1024 & 0.1050 & \(1.14785 \times 10^{-2}\) & 0.0037 & \(1.82629 \times 10^{-5}\) \\ 
Goldwasser-Micali & 1024 & 0.0400 & \(2.74090 \times 10^{-4}\) & 0.0093 & \(6.70433 \times 10^{-5}\) \\ 
Exponential-ElGamal & 1024 & 0.0302 & \(1.09401 \times 10^{-3}\) & 3.1553 & \(1.03474 \times 10^{-5}\) \\ 
EllipticCurve-ElGamal & 160 & 0.0053 & \(1.01023 \times 10^{-2}\) & 4.1529 & \(6.82354 \times 10^{-5}\) \\
\bottomrule
\end{tabular}
\label{tab:performance_80}
\end{table}
\end{minipage}
\newline
\newline

\begin{minipage}{\linewidth} 
\begin{table}[H]
\caption{Time Consumption of Algorithms with 112-bit Symmetric Key Size Equivalent in Seconds}
\centering
\begin{tabular}{llllll}
\toprule
Algorithm & Key Size & Key Generation & Encrypt & Decrypt & Homomorphic Operation \\
\midrule
RSA & 2048 & 1.4959 & \(2.14737 \times 10^{-2}\) & 0.0235 & \(5.04017 \times 10^{-5}\) \\ 
ElGamal & 2048 & 0.6238 & \(7.15775 \times 10^{-3}\) & 0.0037 & \(1.92165 \times 10^{-5}\) \\ 
Paillier & 2048 & 0.7613 & \(8.51427 \times 10^{-2}\) & 0.0847 & \(5.97000 \times 10^{-5}\) \\ 
Damgard-Jurik & 2048 & 0.6388 & \(1.72729 \times 10^{-1}\) & 0.1997 & \(1.04141 \times 10^{-4}\) \\ 
Okamoto-Uchiyama & 2048 & 0.7543 & \(7.80529 \times 10^{-2}\) & 0.0239 & \(3.19004 \times 10^{-5}\) \\ 
Goldwasser-Micali & 2048 & 0.6725 & \(6.73010 \times 10^{-4}\) & 0.0536 & \(1.94454 \times 10^{-4}\) \\ 
Exponential-ElGamal & 2048 & 0.3957 & \(6.95195 \times 10^{-3}\) & 10.572 & \(1.69277 \times 10^{-5}\) \\ 
EllipticCurve-ElGamal & 224 & 0.0060 & \(8.61859 \times 10^{-3}\) & 3.6356 & \(5.64575 \times 10^{-5}\) \\
\bottomrule
\end{tabular}
\label{tab:performance_112}
\end{table}
\end{minipage}
\newline
\newline

\begin{minipage}{\linewidth} 
\begin{table}[H]
\centering
\caption{Time Consumption of Algorithms with 128-bit Symmetric Key Size Equivalent in Seconds}
\begin{tabular}{llllll}
\toprule
Algorithm & Key Size & Key Generation & Encrypt & Decrypt & Homomorphic Operation \\
\midrule
RSA & 3072 & 5.8871 & $0.0886 \times 10^{-2}$ & 0.1201 & $4.106 \times 10^{-5}$ \\
ElGamal & 3072 & 1.5976 & $0.0253 \times 10^{-2}$ & 0.0120 & $3.009 \times 10^{-5}$ \\
Paillier & 3072 & 2.5762 & $0.3211 \times 10^{-1}$ & 0.3188 & $1.361 \times 10^{-4}$ \\
Damgard-Jurik & 3072 & 3.0325 & $0.6973 \times 10^{-1}$ & 0.6809 & $2.054 \times 10^{-4}$ \\
Okamoto-Uchiyama & 3072 & 2.7635 & $0.2815 \times 10^{-1}$ & 0.0881 & $6.485 \times 10^{-5}$ \\
Goldwasser-Micali & 3072 & 3.6569 & $0.0011 \times 10^{-2}$ & 0.4216 & $3.805 \times 10^{-4}$ \\
Exponential-ElGamal & 3072 & 1.6232 & $0.0214 \times 10^{-2}$ & 23.171 & $2.503 \times 10^{-5}$ \\
EllipticCurve-ElGamal & 256 & 0.0086 & $0.0119 \times 10^{-2}$ & 4.3759 & $7.157 \times 10^{-5}$ \\
\bottomrule
\end{tabular}
\label{tab:performance_128}
\end{table}
\end{minipage}
\newline
\newline

\begin{minipage}{\linewidth} 
\begin{table}[H]
\centering
\caption{Time Consumption of Algorithms with 192-bit Symmetric Key Size Equivalent in Seconds}
\begin{tabular}{llllll}
\toprule
Algorithm & Key Size & Key Generation & Encrypt & Decrypt & Homomorphic Operation \\
\midrule
RSA & 7680 & 370.568 & $0.8109 \times 10^{0}$ & 0.9861 & $1.1673 \times 10^{-4}$ \\
ElGamal & 7680 & 22.1267 & $0.2702 \times 10^{0}$ & 0.1383 & $7.1764 \times 10^{-5}$ \\
Paillier & 7680 & 71.0874 & $3.9756 \times 10^{0}$ & 4.0049 & $5.4407 \times 10^{-4}$ \\
Damgard-Jurik & 7680 & 110.567 & $8.0903 \times 10^{0}$ & 8.1464 & $1.0164 \times 10^{-3}$ \\
Okamoto-Uchiyama & 7680 & 84.0547 & $3.8247 \times 10^{0}$ & 1.1667 & $2.6650 \times 10^{-4}$ \\
Goldwasser-Micali & 7680 & 78.9467 & $5.8275 \times 10^{-3}$ & 4.1952 & $2.1928 \times 10^{-3}$ \\
Exponential-ElGamal & 7680 & 49.9941 & $0.2669 \times 10^{0}$ & 117.68 & $7.1144 \times 10^{-5}$ \\
EllipticCurve-ElGamal & 384 & 0.01000 & $7.1723 \times 10^{-3}$ & 3.5553 & $5.3978 \times 10^{-5}$ \\
\bottomrule
\end{tabular}
\label{tab:performance_192}
\end{table}
\end{minipage}
\newline
\newline

\subsection{Cloud Performances}
\label{sec:cloud}

We conducted experiments on key generation, encryption, decryption, and homomorphic operations using various algorithms with an equivalent 192-bit symmetric key size. These experiments were performed across several cloud environments, including Colab Normal, Colab A100 GPU, Colab L4 GPU, Colab T4 High RAM, Colab TPU2, and Azure Spark, running five experiments for each task and averaging the consumption times to ensure accuracy.

We utilize radar map charts to present the performance of various algorithms across diverse cloud environments. Multiple radar maps are shown because consolidating them into a single radar map would render it illegible. Additionally, the radar maps are arranged from left to right to illustrate slower to faster speeds.

Figure \ref{fig:key_generation} illustrates a radar map depicting key generation times in seconds for different algorithms across various cloud environments. Among these algorithms, RSA emerges as the slowest in key generation, with Elliptic Curve ElGamal demonstrating the fastest performance. Following Elliptic Curve ElGamal, ElGamal and Exponential ElGamal exhibit slightly slower key generation times than Elliptic Curve ElGamal. The remaining algorithms exhibit relatively similar performance to each other. Notably, the key generation process exhibits significant instability compared to encryption, decryption, or homomorphic operations. This instability arises from the inherent requirement of generating random values until a specific condition is met, contributing to fluctuating performance levels.

Figure \ref{fig:encrypt} shows the performance comparisons of various algorithms in seconds for encryption tasks across different cloud environments, with values depicted in seconds. Damgard-Jurik, Paillier, and Okamoto-Uchiyama were the slowest performers. ElGamal and Exponential ElGamal charts are overlaid, showcasing moderate performance, with RSA slightly trailing behind in this category. Meanwhile, Elliptic Curve ElGamal emerged as the fastest option.

The decryption performance comparison in seconds, shown in Figure \ref{fig:decrypt}, revealed Exponential ElGamal as the slowest, ElGamal as the fastest, and other algorithms displaying moderate performance across different cloud environments.

Figure \ref{fig:homomorphic} highlights the performance of various algorithms in seconds for homomorphic operations across different cloud environments, with values depicted in seconds. Damgard-Jurik, Goldwasser-Micali, and Paillier were the slowest, while ElGamal, Exponential ElGamal, and Elliptic Curve ElGamal exhibited the fastest performance. Okamoto-Uchiyama and RSA demonstrated moderate performance in homomorphic operations. In practice, just homomorphic opetaion will be performed on the cloud environment and just this operation will be done often.

In evaluating the performance of various tasks across different cloud environments, it becomes evident that the Colab Normal environment emerges as the slowest, contrasting sharply with the remarkable speed demonstrated by Colab TPU2 and Colab TPU High RAM, irrespective of the algorithm employed. The performances of various tasks across different cloud environments are summarized in Tables \ref{tab:evaluate_cloud_part1} and \ref{tab:evaluate_cloud_part2}.

\begin{minipage}{\linewidth} 
\begin{table}[H]
\centering
\caption{Cloud Environment Configurations}
\begin{tabular}{llll}
\toprule
Environment & Processor & Resource & Memory \\
\midrule
Colab CPU & CPU & Intel Xeon CPU & $\sim$13 GB \\
Colab A100 GPU & GPU & NVIDIA A100 & 40 GB HBM2 \\
Colab TPU2 & TPU &  Tensor Processing Unit & $\sim$16 GB HBM \\
Colab L4 GPU & GPU & NVIDIA L4 & 24 GB GDDR6 \\
Colab T4 High RAM & GPU & NVIDIA T4 & 32 GB GDDR6 \\
Azure Spark &  Distributed System &  Varies by configuration & Varies \\
\bottomrule
\end{tabular}
\label{tab:evaluate_cloud_part1}
\end{table}
\end{minipage}
\newline
\newline

\begin{minipage}{\linewidth} 
\begin{table}[H]
\footnotesize
\centering
\caption{Evaluation of Cloud Performances}
\begin{tabular}{lllll}
\toprule
Environment & Key Generation & Encrypt & Decrypt & Homomorphic Operations \\
\midrule
Colab CPU & Moderate & Moderate & Moderate & Moderate, limited by single-thread processing \\
Colab A100 GPU & Fast & Fast & Fast & Very Fast, optimized for parallel processing \\
Colab TPU2 & Fast & Fast & Fast & Fast, optimized for tensor operations \\
Colab L4 GPU & Fast & Fast & Fast & Very Fast, suitable for a variety of workloads \\
Colab T4 High RAM & Fast & Fast & Fast & Fast, enhanced for memory-intensive tasks \\
Azure Spark & Mod to Fast & Mod to Fast & Mod to Fast & Mod to Fast, depending on node configuration \\
\bottomrule
\end{tabular}
\label{tab:evaluate_cloud_part2}
\end{table}
\end{minipage}
\newline
\newline

\begin{minipage}{\linewidth} 
\begin{figure}[H]
    \centering
    \includegraphics[width=0.99\textwidth]{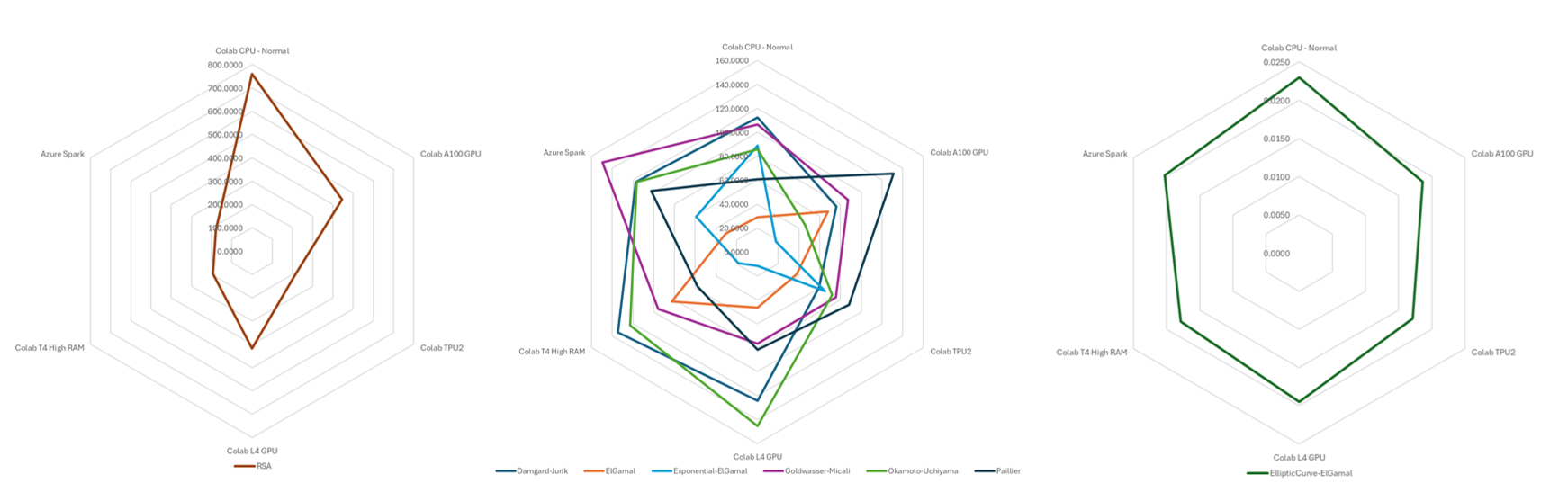}
    \caption{Key Generation Radar Map of Cloud Environments With Respect To The Algorithms}
    \label{fig:key_generation}
\end{figure}
\end{minipage}

\begin{minipage}{\linewidth} 
\begin{figure}[H]
    \centering
    \includegraphics[width=0.99\textwidth]{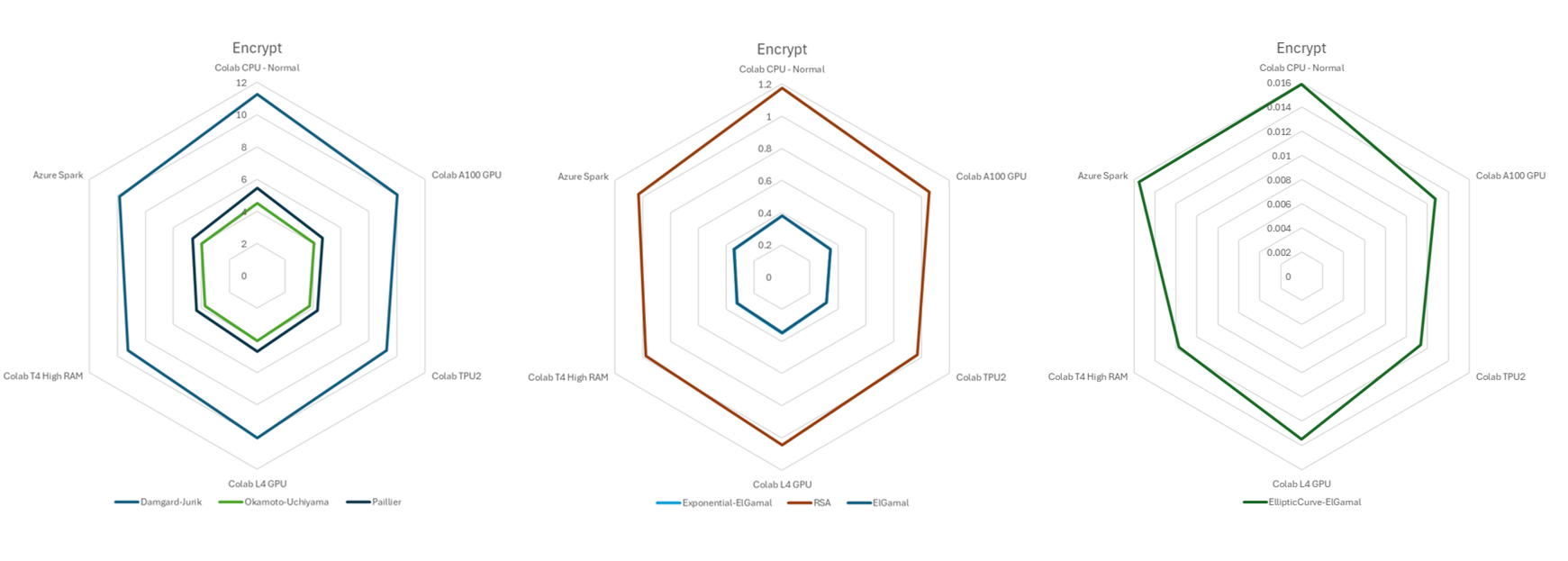}
    \caption{Encryption Performances of Algorithms for Different Cloud Environments}
    \label{fig:encrypt}
\end{figure}
\end{minipage}

\begin{minipage}{\linewidth} 
\begin{figure}[H]
    \centering
    \includegraphics[width=0.99\textwidth]{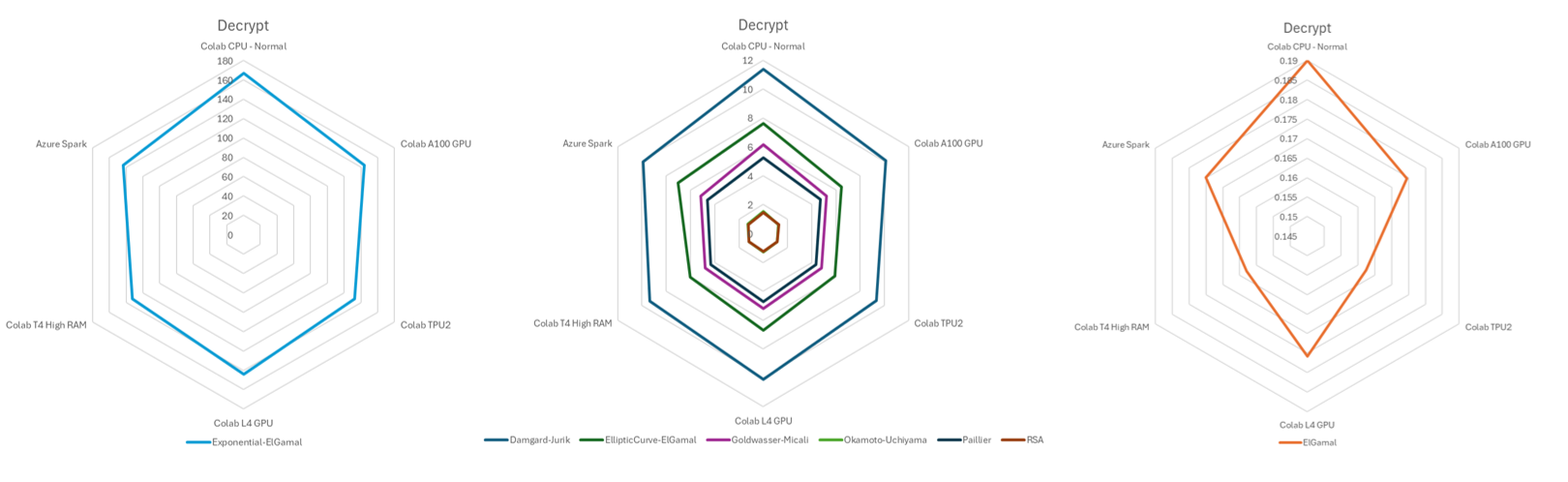}
    \caption{Decryption Performances of Algorithms for Different Cloud Environments}
    \label{fig:decrypt}
\end{figure}
\end{minipage}

\begin{minipage}{\linewidth} 
\begin{figure}[H]
    \centering
    \includegraphics[width=0.99\textwidth]{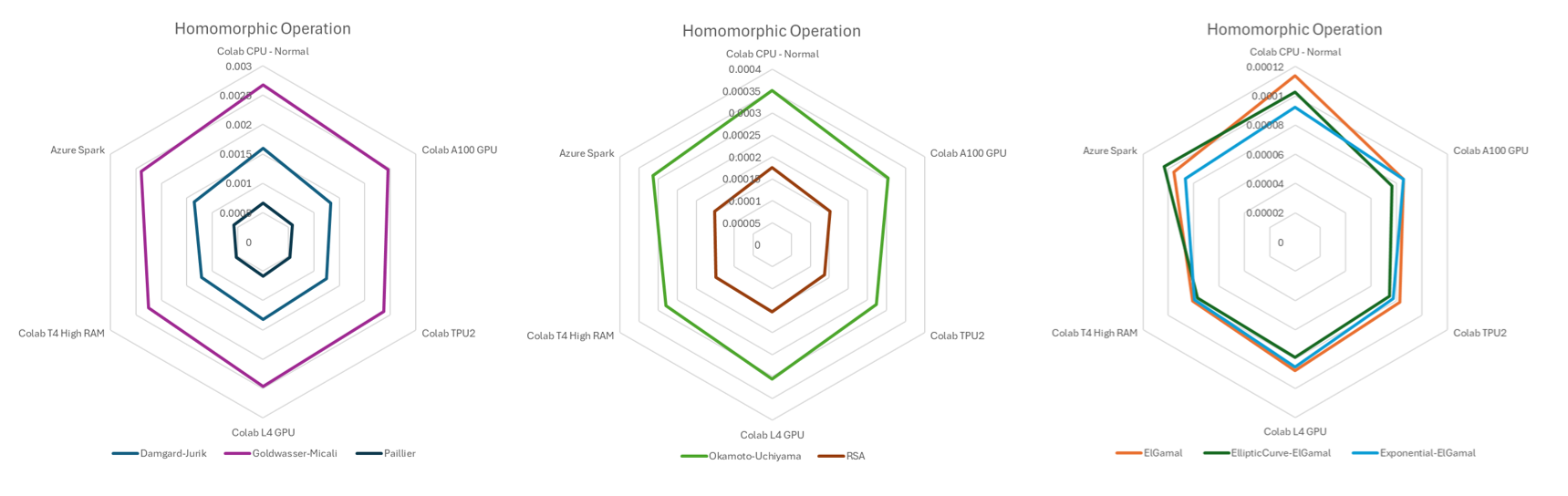}
    \caption{Homomorphic Operation Performances of Algorithms for Different Cloud Environments}
    \label{fig:homomorphic}
\end{figure}
\end{minipage}

\section{Conclusion}
\label{sec:conclusion}

In conclusion, the evolution of encryption technologies has reached a pivotal point. This paper introduces LightPHE, a novel framework for partially homomorphic encryption (PHE) designed to streamline cryptographic operations in practical applications. LightPHE serves as a unified interface that encapsulates multiple PHE algorithms, providing developers with a flexible and efficient toolset for securing sensitive data while preserving computational functionality.

The exploration of LightPHE commenced with an in-depth examination of the underlying mathematical principles behind various PHE algorithms, illuminating their homomorphic features and capabilities. By elucidating these mathematical foundations, the groundwork was laid for understanding how LightPHE facilitates secure and efficient computation on encrypted data.

The software design and architecture of LightPHE were then detailed, emphasizing its modular structure and abstract class hierarchy. This architectural design enables seamless integration of new PHE algorithms while maintaining a consistent interface for users. The elegant design of LightPHE promotes code reusability and extensibility, facilitating rapid prototyping and development of secure applications.

To demonstrate the practical utility of LightPHE, detailed code snippets were provided, illustrating its implementation in real-world scenarios. These examples showcased how developers can leverage LightPHE to perform homomorphic addition, multiplication, and other cryptographic operations with ease. By abstracting away the complexities of individual PHE algorithms, LightPHE empowers developers to focus on building secure and efficient applications without being burdened by cryptographic intricacies.

Extensive experiments were conducted to evaluate the performance of LightPHE across different key sizes and various cloud environments, including Colab Normal, Colab A100 GPU, Colab L4 GPU, Colab T4 High RAM, Colab TPU2, and Azure Spark cloud. These experiments provided valuable benchmarks for understanding the efficiency and scalability of LightPHE in diverse computing settings. The results highlighted the practical applicability of LightPHE, demonstrating its performance in terms of key generation, encryption, decryption, and homomorphic operations across different cloud configurations.

The cloud environment benchmarks illustrated that LightPHE performs optimally in high-performance settings such as Colab A100 GPU and TPU2, which are optimized for parallel processing and tensor operations. These environments showed superior computational power, making them suitable for tasks requiring high performance and scalability. Conversely, more cost-effective and accessible options like Colab Normal and Azure Spark offered viable alternatives for projects with limited resources.

Ultimately, LightPHE represents a significant advancement in the field of partially homomorphic encryption, offering a versatile and user-friendly framework for implementing secure computation protocols. By bridging the gap between theory and practice, LightPHE paves the way for the widespread adoption of homomorphic encryption in diverse application domains, from healthcare to finance and beyond. Future research endeavors may explore optimizations and enhancements to further improve the efficiency and scalability of LightPHE, unlocking new possibilities for secure and privacy-preserving computation in the digital age.

\bibliographystyle{unsrt}
\bibliography{template}

\end{document}